\newcommand{\atlas}{ATLAS$^{\rm 3D}$}
\newcommand{\hi}{{\sc H\,i}}
\newcommand{\mhi}{$M$(\hi)}

\newcommand{\msun}{M$_\odot$}

\documentclass[cits]{PoS}

\title{An \hi\ view of the on-going assembly of early-type galaxies: present and future observations}

\ShortTitle{An \hi\ view of the on-going assembly of early-type galaxies}

\author{\speaker{P.~Serra}\\
        ASTRON, Postbus 2, 7990 AA Dwingeloo, the Netherlands\\
        E-mail: \email{serra@astron.nl}}
        
\author{R.~Morganti$^{1,2}$, T.~A.~Oosterloo$^{1,2}$, K.~Alatalo$^{3}$, L.~Blitz$^{3}$, M.~Bois$^{4}$, R.~C.~E.~van~den~Bosch$^{5}$, F.~Bournaud$^{6}$, M.~Bureau$^{7}$, M.~Cappellari$^{7}$, R.~L.~Davies$^{7}$, T.~A.~Davis$^{7}$, P.~Duc$^{6}$, E.~Emsellem$^{4,8}$, J.~Falc\'{o}n-Barroso$^{10}$, S.~Khochfar$^{11}$, D.~Krajnovi\'{c}$^{8}$, H.~Kuntschner$^{8}$, P.-Y.~Lablanche$^{4}$, R.~M.~McDermid$^{12}$, T.~Naab$^{13}$, M.~Sarzi$^{14}$, N.~Scott$^{7}$, G.~van~de~Ven$^{15}$, A.~Weijmans$^{16}$, L.~M.~Young$^{17}$ and P.~T.~de~Zeeuw$^{8,9}$\\
$^1$ASTRON, Dwingeloo, the Netherlands, $^2$Kapteyn Astronomical Institute, Groningen , the Netherlands, 
$^3$University of California, Berkeley, USA, 
$^4$Universit\'e de Lyon, France, 
$^5$University of Texas, Austin, USA, 
$^6$Universit\'e Paris Diderot, France, 
$^7$University of Oxford, UK, 
$^8$European Southern Observatory, Garching, Germany, 
$^9$Leiden Observatory, the Netherlands, 
$^{10}$Instituto de Astrof\'isica de Canarias, La Laguna, Spain, 
$^{11}$Max-Planck-Institute for Extraterrestrial Physics, Garching, Germany, 
$^{12}$Gemini Observatory, Hilo, USA, 
$^{13}$Universit\"{a}ts-Sternwarte M\"{u}nchen, Germany, 
$^{14}$University of Hertfordshire, Hatfield, UK, 
$^{15}$Max Planck Institute for Astronomy, Heidelberg, Germany
$^{16}$Dunlap Institute for Astronomy and Astrophysics, Toronto, Canada, 
$^{17}$New Mexico Tech, Socorro, USA
}

\abstract{

We present a preliminary analysis of the \hi\ properties of early-type galaxies in the \atlas\ sample. Using WSRT data for $\sim$100 galaxies outside the Virgo cluster and data from the Alfalfa project for galaxies inside Virgo, we discuss the dependence of \hi\ properties on environment. We detect \hi\ in about half of the galaxies outside Virgo. For these systems, the \hi\ morphology and kinematics change as a function of environment, going from regular, rotating systems around ``isolated'' galaxies to progressively more disturbed structures for galaxies with neighbours or in groups. In denser environment, inside Virgo, nearly none of the galaxies contains \hi.

We discuss future work in this field which will be enabled by next-generation, pre-SKA radio instruments. We present a simulated Apertif \hi\ observation of an \atlas\ early-type galaxy, showing how its appearance and detection level vary as a function of redshift.}

\FullConference{Panoramic Radio Astronomy: Wide-field 1-2 GHz research on galaxy evolution - PRA2009\\
		 June 02 - 05 2009\\
		 Groningen, the Netherlands}

\begin{document}

\section{Early-type galaxies and the \atlas\ project}
\label{sec1}

Nearby early-type galaxies (ETGs) are a fundamental class of objects. They provide a fossil record of galaxy assembly and allow us to study the physical processes that drive the continuing evolution of galaxies at $z$$\sim$0. Their proximity enables detailed studies at high spatial/spectral resolution and high signal-to-noise ratio. However, work done so far has been limited to relatively small and incomplete  samples (e.g., \cite{2002MNRAS.329..513D,2005ApJ...621..673T}). The \atlas\ project, which we discuss in this contribution, is an attempt to overcome this limitation.

\atlas\ (http://purl.org/atlas3d) is a complete, multi-wavelength, volume-limited survey of 262 nearby ETGs, supported by numerical simulations and semi-analytic models of galaxy assembly. The \atlas\ sample is selected on the basis of optical morphology from a parent sample of 844 galaxies (of all types) meeting the criterions distance$\leq$42 Mpc and M$_{\rm K}\leq -21.5$ (with additional constraints on declination and galactic latitude). The morphological classification is based on the (non-) detection of spiral arms in optical images taken from SDSS DR6 (available for 82\% of the parent-sample) or DSS2-blue. A detailed description of the sample selection and its characterisation in terms of, for example, luminosity function, distribution on the colour-magnitude diagram and environment can be found in Cappellari et al.\ (in prep.). Nearly all \atlas\ galaxies lie on the bright part of the red sequence defined in \cite{2004ApJ...600..681B}. The Virgo cluster of galaxies is included in the survey, so that the sample spans a factor of $\sim10^3$ in environment density. The luminosity function of the\atlas\ sample follows that of larger, $z$$\sim$0 samples (e.g., \cite{2003ApJS..149..289B}).

Data taken within the \atlas\ project include optical integral-field spectroscopy with the SAURON spectrograph; molecular-gas, millimetre, single-dish observations with IRAM and interferometric follow-up of detections with CARMA; and neutral-hydrogen 21-cm interferometry with the WSRT. Furthermore, INT SDSS-like imaging has been obtained for \atlas\ galaxies outside the SDSS coverage.

The observation of \hi\ gas in ETGs is a fundamental component of the \atlas\ project. In particular, deep radio interferometry allows us to determine the detailed morphology and kinematics of the \hi\ around ETGs. As proven by previous work, this is important to study the state of the cold gas as a signature of the recent gas-accretion history of the host galaxy (\cite{2006MNRAS.371..157M}). In this contribution, we present preliminary results of our \hi\ observations focusing on the dependence of ETG \hi\ properties on environment.

\section{HI and environment of early-type galaxies}

\begin{figure}
\centering
\includegraphics[width=9cm]{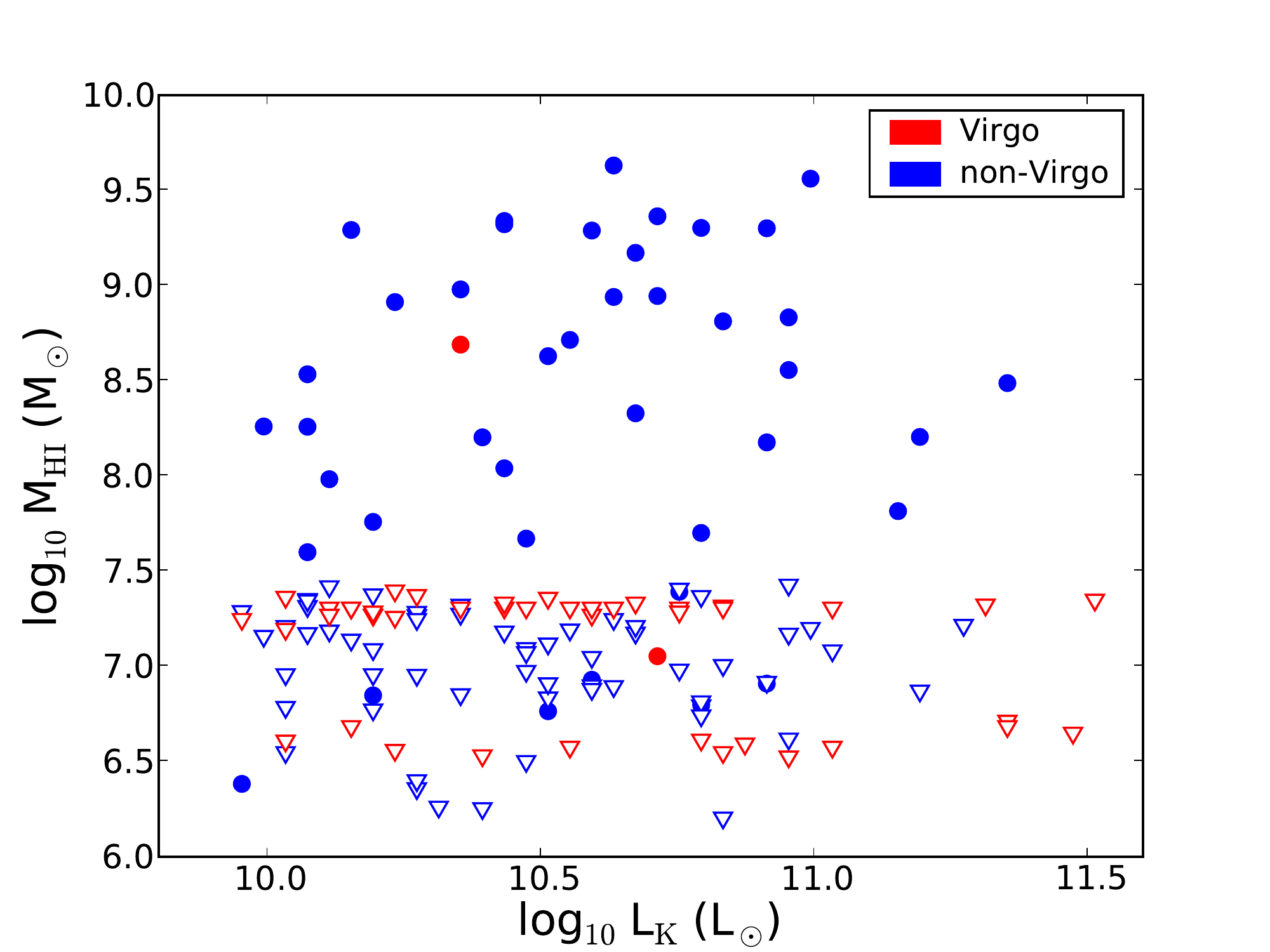}
\caption{\hi\ content of nearly all ETGs in the \atlas\ sample at DEC$\geq$+10 deg. Red and bue symbols correspond to Virgo and non-Virgo galaxies respectively. Filled circles are \hi-detected galaxies, open triangles are upper limits on \mhi\ (see text).}
\label{fig1}
\end{figure}

The \atlas\ \hi\ observations consist of a 12-h WSRT integration per galaxy for objects at $\delta\geq$+10 deg and outside the Virgo cluster. Galaxies inside the Virgo cluster have been observed by the Alfalfa collaboration (\cite{2005AJ....130.2598G,2007A&A...474..851D}) and were re-observed by us with the WSRT only in the few cases of Alfalfa detection. In Fig.\ref{fig1} we show the detected \hi\ mass plotted against the $K$-band total luminosity for nearly all objects. Red and blue symbols represent galaxies inside and outside the Virgo cluster respectively. Filled circles and open triangles represent detected and non-detected objects respectively. In case of undetected objects, the upper limit on \mhi\ is calculated assuming a 3$\sigma$ signal spread over 200 km/s. Outside Virgo, about half of all ETGs contain \hi\ at the \mhi$\geq$10$^7$~\msun\ level. This confirms earlier (but less complete) results showing that \hi\ is a fairly common characteristic of ETGs outside the densest environments (\cite{2006MNRAS.371..157M,2007A&A...465..787O,2009A&A...498..407G}).

The distribution of points in Fig.\ref{fig1} shows a very strong environmental effect: ETGs become much less likely to contain \hi\ if they reside inside the Virgo cluster (see \cite{2007A&A...474..851D,2009A&A...498..407G}). The hot intra-cluster medium and the high galaxy-density in Virgo prevent galaxies from retaining and/or accreting \hi, as is evident also from the observation of late-type objects by \cite{2009arXiv0909.0781C}.

Our WSRT \hi\ images reveal another interesting environmental effect for galaxies outside the Virgo cluster. Gas in/around galaxies residing in the lowest-density environment appears more relaxed than gas in/around galaxies residing in, e.g., groups. To perform this analysis we define the environment of a galaxy as a cylinder of radius 2 Mpc on the sky and depth 600 km/s in line-of-sight velocity, centred on the galaxy itself. The total luminosity $L_\mathrm{env}$ of galaxies inside this cylinder was calculated using the parent sample described in Sec.\ref{sec1}, limited to M$_{\rm K}\leq -21.5$. We then adopt the ratio $r$=$L_\mathrm{galaxy}$/$L_\mathrm{env}$ to characterise the level of isolation of a given galaxy in \atlas. Galaxies with $r\sim1$ dominate their environment and may be surrounded by just a few much smaller satellites. Galaxies with $r=0.01$-0.1 reside in rich environments and are close to many objects of similar or larger mass.

\begin{figure}
\centering
\includegraphics[width=15cm]{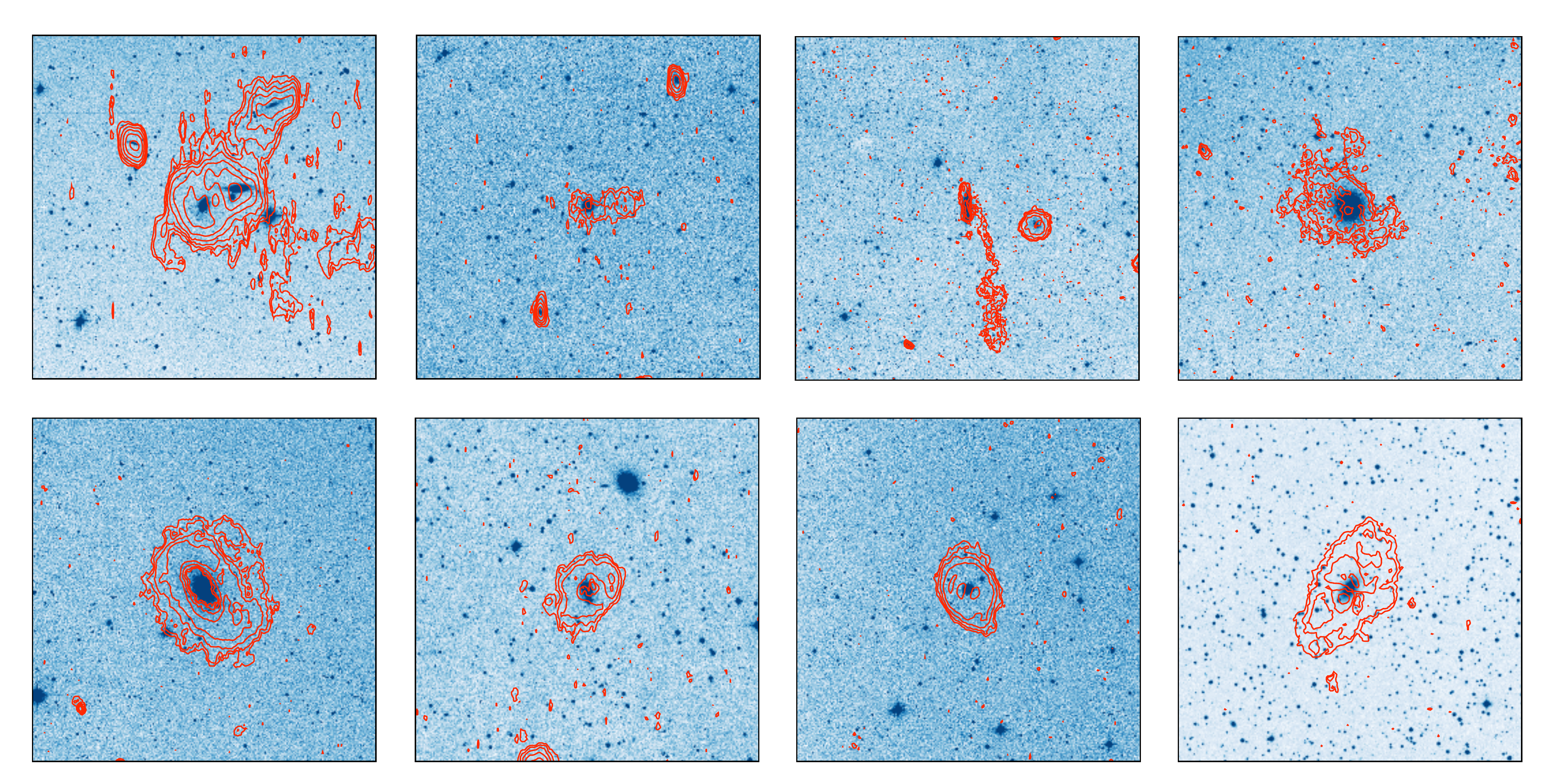}
\caption{Preliminary \hi\ images of some of the galaxies in the \atlas\ sample outside the Virgo cluster. Red contours of constant \hi\ column density are plotted on top of optical DSS images. All images have a side of 180 kpc. \hi\ contours are drawn at $\sim$1.5$\cdot$2$^n$$\times$10$^{19}$ cm$^{-2}$ ($n$=0,1,2,...). The top row shows galaxies in higher-density environment (e.g., groups). The bottom row shows isolated galaxies (see text for our definition of isolation).}
\label{fig2}
\end{figure}

In Fig.\ref{fig2} we show a few typical examples of \hi-detected \atlas\ ETGs outside the Virgo cluster. The top row contains galaxies in higher-density environment (e.g., small groups), where $r$ is close to zero. The bottom row contains ``isolated'' galaxies, i.e., systems for which no other galaxy could be found within the cylinder defined above (and at the magnitude limit of the parent sample) and therefore $r=1$. The \hi\ around ``isolated'' galaxies is typically found in extended, regularly-rotating distributions, significantly more relaxed than the \hi\ detected around objects in denser environments.

This analysis can be and has been quantified by measuring the asymmetry of the \hi\ distribution on the sky for each galaxy. Even better, asymmetry can be measured in the 3-dimensional [$\alpha$,$\delta$,$v_\mathrm{l.o.s.}$] space (an ideal, rotating disc/ring is symmetric with respect to the origin in this space), and such quantitative approach is a promising technique to apply to future, large \hi\ surveys (see below). Although we do not enter in the detail of our preliminary analysis in this contribution, we can say that it confirms the visual impression given by Fig.\ref{fig2}.

The rate at which galaxies interact with their satellites and neighbours is naturally determined by the environment density. The \hi\ properties of ETGs displayed in Figs.\ref{fig1} and \ref{fig2} show convincingly that such interaction is still occurring at $z$$\sim$0 and is a fundamental driver of ETG gas properties. In the most quiet environment, there is sufficient time for ETGs to accrete regular but dilute \hi\ systems that can survive for many Gyrs (i.e., many gas orbits) without forming a significant stellar disc.

\section{Future work: simulated Apertif observation of early-type galaxies}

The coming decades will see a dramatic improvement in the number and survey-speed of radio telescopes. Instruments like Apertif (\cite{PoS(PRA2009)006}) and ASKAP (\cite{PoS(PRA2009)002}) will revolutionise radio astronomy by enabling all-sky line and continuum surveys at high spatial and spectral resolution and with a wide bandpass (i.e., covering a large redshift range at once). One crucial aspect when planning an \hi\ survey is understanding to what redshift \hi\ will be detectable, and what kind of information it will be possible to extract from the data cubes. With respect to the observation of ETGs, the \atlas\ dataset represent a precursor of the future surveys. The \atlas\ \hi\ dataset is a unique opportunity to perform realistic simulations of what we will see with the new telescopes.

With this in mind, we used the WSRT datacube of one particular galaxy belonging to the \atlas\ sample to simulate a 10$\times$12-h Apertif observation as the galaxy moves to higher redshift. To do so, we make use of the \hi\ CLEAN model resulting from the deconvolution performed during the data reduction. For a number of values of the redshift $z$, we reduce the size and flux of the CLEAN model by re-scaling it to the corresponding angular-size distance and luminosity distance respectively. We then convolve the redshifted model with the synthesised beam used for the original data. Finally, we add to it a noise cube also derived from the original 12-h WSRT data and scaled down by a factor $\sqrt{10}$ (current noise estimate for Apertif give the same noise as that of recent WSRT data).

\begin{figure}
\centering
\includegraphics[width=15cm]{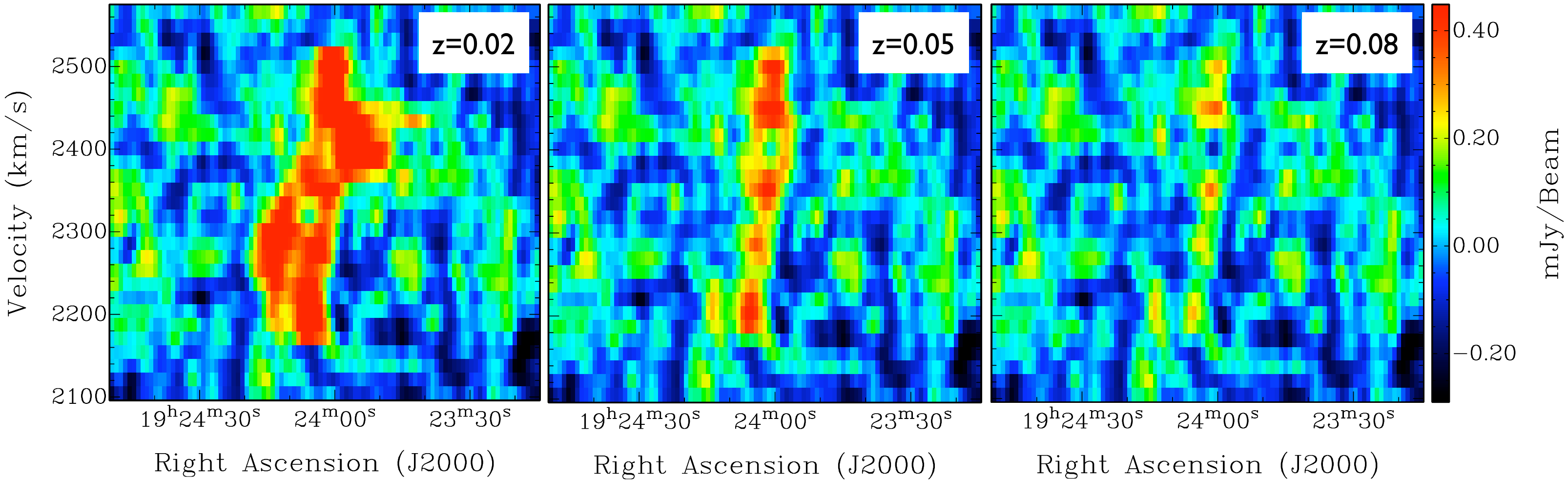}
\caption{Position-velocity diagrams extracted from the 10$\times$12-h simulated Apertif observation of a galaxy in the \atlas\ sample. The three panels correspond to a galaxy redshift $z=0.02,0.05,0.08$ from left to right. See text for a description of the simulation.}
\label{fig3}
\end{figure}

The result of this simulation is shown in Fig.\ref{fig3}. While the \hi\ disc surrounding the galaxy is still clearly visible at $z=0.05$, it is barely detected at $z=0.08$ and very little information other than the total \hi\ mass could be derived from the data at this redhift. Of course, the \hi\ observation of ETGs is particularly challenging because of the typically low column-density of the gas. For example, a late-type galaxy of similar mass (in this case, 2.3$\times$10$^9$ \msun) should be detectable up to higher redshift. This simulation is therefore useful to understand to what redshift it is possible to push work similar to the one presented in the previous sections with the large surveys planned for the coming years.

\section{Conclusions}

We have presented preliminary results of an \hi\ WSRT survey of ETGs in the \atlas\ sample. We confirm a strong dichotomy between Virgo and non-Virgo galaxies, with the former being systematically \hi-poorer than the latter. We find that, outside Virgo, \hi\ systems become progressively more settled as the environment density decreases. ``Isolated'' ETGs are characterised by very regular, rotating cold-gas systems while galaxies with neighbours or residing in galaxy groups typically have disturbed \hi\ morphology/kinematics. This confirms that environment is a fundamental driver of ETG's evolution at $z$$\sim$0 and that, in the absence of significant disturbance from nearby objects, ETGs can accrete and maintain large systems of cold gas without this appreciably affecting their optical morphology.

We have also presented a simulated Apertif observation of an ETG belonging to the \atlas\ sample as a function of the galaxy's redshift. With an integration of 10$\times$12 h, the low-column-density $\sim2\times10^9$ \msun\ of \hi\ surrounding the galaxy remains detectable (and the gas distribution/kinematics can be studied in detail) out to $z\sim0.05$. However, by $z\sim0.1$ the \hi\ is lost to the noise. This simulation is useful to set what a reasonable expectation should be for the \hi\ observation of ETGs using upcoming all-sky 21-cm surveys.


\begin{thebibliography}{99}

\bibitem{2004ApJ...600..681B} I.~K.~Baldry et al., \emph{Quantifying the Bimodal Color-Magnitude Distribution of Galaxies}, 2004, ApJ, 600, 681B
\bibitem{2003ApJS..149..289B}E.~F.~Bell, D.~H.~McIntosh, N.~Katz, M.~D.~Weinberg, \emph{The Optical and Near-Infrared Properties of Galaxies. I. Luminosity and Stellar Mass Functions}, 2003, ApJ, 149, 289B
\bibitem{2009arXiv0909.0781C} A.~Chung, J.~H.~van Gorkom, J.~D.~P.~Kenney, H.~Crowl, B.~Vollmer, \emph{VLA Imaging of Virgo Spirals in Atomic Gas (VIVA). I. The Atlas and the \hi\ Properties}, 2009, AJ, 138, 1741C
\bibitem{PoS(PRA2009)002}I.~Feain, S.~Johnston, R.~Braun, \emph{The Australian Square Kilometre Array Pathfinder (ASKAP) an SKA pre-cursor}, in proceedings of \emph{Panoramic Radio Astronomy 2009}, \pos{PoS(PRA2009)002}
\bibitem{2005AJ....130.2598G}Giovanelli~R. et al., \emph{The Arecibo Legacy Fast ALFA Survey. I. Science Goals, Survey Design, and Strategy} 2005, AJ, 130, 2598G
\bibitem{2009A&A...498..407G}M.~Grossi et al., \emph{The \hi\ content of early-type galaxies from the ALFALFA survey. II. The case of low density environments}, 2009, A\&A, 498, 407
\bibitem{2006MNRAS.371..157M}R.~Morganti et al., \emph{Neutral hydrogen in nearby elliptical and lenticular galaxies: the continuing formation of early-type galaxies}, 2006, MNRAS, 371, 157
\bibitem{2007A&A...465..787O}T.~A.~Oosterloo, R.~Morganti,  E.~M.~Sadler,  J.~M.~van~der~Hulst, P.~Serra, \emph{Extended, regular \hi\ structures around early-type galaxies}, 2007, A\&A, 465, 787
\bibitem{PoS(PRA2009)006}T.~A.~Oosterloo, W.~van~Cappellen, M.~Verheijen, L.~Bakker, G.~Heald \emph{The latest on Apertif}, in proceedings of \emph{Panoramic Radio Astronomy 2009}, \pos{PoS(PRA2009)006}
\bibitem{2007A&A...474..851D}S.~di Serego Alighieri et al., \emph{The HI content of early-type galaxies from the ALFALFA survey. I. Catalogued HI sources in the Virgo cluster}, 2007, A\&A, 474, 851
\bibitem{2005ApJ...621..673T}D.~Thomas, C.~Maraston, R.~Bender, C.~Mendes~de~Oliveira, \emph{The Epochs of Early-Type Galaxy Formation as a Function of Environment}, 2005, ApJ, 621, 673
\bibitem{2002MNRAS.329..513D}P.~T.~de~Zeeuw et al., \emph{The SAURON project - II. Sample and early results}, 2002, MNRAS, 329, 513

\end{thebibliography}
\end{document}